**Transfer of Phase Information between Microwave and Optical Fields via an Electron Spin**

Ignas Lekavicius, D. Andrew Golter, Thein Oo, and Hailin Wang

Department of Physics, University of Oregon, Eugene, OR 97403

**Abstract**

We demonstrate the coherent coupling and the resulting transfer of phase information between microwave and optical fields in a single nitrogen vacancy center in diamond. The relative phase of two microwave fields is encoded in a coherent superposition spin state. This phase information is then retrieved with a pair of optical fields. A related process is also used for the transfer of phase information from optical to microwave fields. These studies show the essential role of dark states, including optical pumping into the dark states, in the coherent microwave-optical coupling and open the door to the full quantum state transfer between microwave and optical fields in a solid-state spin ensemble.

Advances in quantum information processing using superconducting circuits have stimulated strong interest in developing interfaces between microwave and optical fields[1,2]. While microwave fields can enable on-chip or local communications between superconducting qubits, optical fields remain the ideal choice for off-chip or long distance communications. Interfaces between microwave and optical fields via their coupling to a spin ensemble in solids or cold atoms have been proposed [3-9], though cold atoms are typically not compatible with the environment needed for superconducting circuits. Other experimental platforms, such as optomechanical resonators[10-13] and electro-optical modulators[14,15], have also been explored. Optomechanics-based systems are susceptible to thermal mechanical noise. The small nonlinear coefficient for microwave-optical coupling in electro-optical modulators limits the photon conversion efficiency. A spin-based interface using systems such as negatively-charged nitrogen vacancy (NV) centers in diamond avoids these limitations.

Negatively-charged NV centers feature robust electron spin coherence[16]. The special energy level structure of a NV center allows this spin coherence to couple to both microwave and optical fields[17,18], though experimental studies have thus far focused on either coherent microwave or optical interactions. Quantum control of the electron spin states has been carried out via direct microwave transitions[19,20], or via optical Raman transitions[21-23]. The robust spin coherence also enables dark-state based optical processes such as coherent population trapping (CPT) and stimulated Raman adiabatic passage[21,22,24-26]. Electron spin ensembles in diamond as well as rare-earth-ion doped crystals have been coupled to a superconducting resonator or qubit via microwave transitions[27-35]. This unique combination of optical and microwave properties suggests that electron spin coherence in diamond can provide an excellent experimental platform to mediate coherent coupling and to develop an interface between optical and microwave fields.

In this paper, we demonstrate that a single NV center in diamond can enable the coherent coupling and especially the transfer of phase information between microwave and optical fields. We map the relative phase of two microwave fields to a superposition spin state in a NV center and then read out the phase with two optical fields. A closely-related process is used for the transfer of phase information from optical to microwave fields. We show that a special superposition spin state, the so-called dark-state, which is decoupled from either optical or microwave fields, plays a central role in the coherent microwave-optical coupling. Similar coherent coupling in a NV ensemble can enable the full quantum-state transfer between microwave and optical fields. In addition, our studies also reveal the important effect of optical pumping on the preparation and readout of superposition spin states. The optical pumping hinders optical readout of the superposition states, but facilitates optical preparation of the dark state as well as the mapping of optical phase information to the dark state.

Our scheme for coherent coupling between optical and microwave fields exploits coherent superposition of the $m_s = \pm 1$ ground spin states in a NV center. As illustrated in Fig. 1, this superposition can couple to a pair of microwave fields via the transitions to the $m_s$=0 ground spin state and to a pair of optical fields via the electric dipole transitions to an excited state, such as the $A_2$ state. To demonstrate that this energy level structure can enable the interfacing between optical and microwave fields, we first map or encode the relative phase of the two microwave fields into the relative phase, $\theta$, of a superposition spin state,

$$| \psi > = e^{i\theta} | C_+ \| +1 > + | C_- \| -1 > \qquad (1)$$

where $C_\pm$ are the complex probability amplitudes for states $|\pm 1 >$ (see Fig. 1a). This relative phase is then read out by the two optical fields (see Fig. 1b). Similarly, encoding the relative optical phase into the superposition spin state and then retrieving the phase information with a pair of microwave fields can transfer phase information from optical to microwave fields.

An electronic grade single-crystal diamond is used in our experiment. The sample is cooled to 10 K in a closed-cycle optical cryostat. Fluorescence from a native NV center about 5 μm below the surface is collected with a confocal microscopy setup. A permanent magnet splits the two ground-state spin transitions by approximately 200 MHz. Time-resolved fluorescence is performed with an avalanche photodiode with a time resolution of 2.8 ns. Details on the phase control of optical and microwave fields as well as the experimental setup are presented in the Supplement[36]. Unless otherwise specified, the Rabi frequency of the microwave fields used is 0.91 MHz, which is small compared with the NV hyperfine splitting (2.2 MHz) such that a microwave field can couple resonantly to only one set of hyperfine states. The corresponding duration for π/2 and π pulse is 225 ns and 550 ns, respectively. The optical Rabi frequency used is 27 MHz, determined from power broadening of the dipole optical transition[37].

For the transfer of phase information from microwave to optical fields, we first initialize the NV center in the $m_s$=0 state and then apply a resonant microwave π/2 pulse to create a superposition of the $m_s$=0 and $m_s$=-1 states (see Fig. 2a). This is followed by a resonant microwave $\pi$ pulse, which excites the remaining electron population in |0> to |+1>, resulting in a superposition spin state given in Eq. 1, with |$C_+$|=|$C_-$|=1/$\sqrt{2}$ and with the relative phase of the superposition state, $\theta$, set by the relative phase of the two microwave fields, $\varphi_{mw} = \varphi_+ - \varphi_-$, where $\varphi_\pm$ are the initial phases of the microwave fields coupling to the $m_s = \pm1$ states [36].

To retrieve the phase information, we apply two optical fields, which have opposite circular polarization and equal Rabi frequency and are resonant with the respective circularly-polarized optical transitions (see Fig. 1b). The two optical fields are Raman resonant with the $m_s = \pm1$ states unless otherwise specified and feature a well-defined relative phase, $\phi_{opt} = \phi_+ - \phi_-$, where $\phi_\pm$ are the initial phases of the optical fields coupling to the $m_s = \pm1$ states. Under these conditions, a superposition spin state given by

$$|D> = (e^{i\phi_{opt}}|+1> - |-1>)/\sqrt{2} \qquad (2a)$$

is effectively decoupled from the optical fields[36]. This state is the optical dark state. Optical excitations to |$A_2$> take place via the bright state,

$$|B> = (e^{i\phi_{opt}}|+1> + |-1>)/\sqrt{2}, \qquad (2b)$$

which is orthogonal to |D>.

The superposition spin state created by the microwave fields can be expressed in the basis of the optical dark and bright states[36],

$$|\psi> = \frac{1}{2}[(e^{i(\theta-\phi_{opt})}+1)|B> + (e^{i(\theta-\phi_{opt})}-1)|D>]. \qquad (3)$$

The optical excitation, i.e. population in |$A_2$>, should scale linearly with the population in |B>, which depends on the relationship between $\theta$ and $\phi_{opt}$, as shown in Eq. 3. The optical readout process, however, can also drive the superposition state into the optical dark state through optical pumping, erasing the phase information encoded by the microwave fields. It is thus essential that the optical readout be performed in a timescale short compared with or at least comparable to the optical pumping time to the dark state (which is of order 30 ns, as will be discussed later).

Figure 2b shows the fluorescence from |$A_2$>, collected during the first 28 ns, as a function of $\phi_{opt}$. The sinusoidal oscillation of the fluorescence in Fig. 2b demonstrates the effective readout of the relative phase of the superposition spin state by the optical fields. Specifically, the

maximum in the oscillation occurs when $\phi_{opt}$ and $\theta$ are in phase. The minimum occurs when $\phi_{opt}$ and $\theta$ are $\pi$ out of phase, for which the electron is initially trapped in the optical dark state. Figure 2c also shows examples of time-resolved fluorescences obtained near a maximum and a minimum of the oscillation. Note that the relative optical (or microwave) phases plotted in Fig. 2 (or Fig. 4) can differ respectively from $\phi_{opt}$ (or $\varphi_{mw}$) by a constant.

The visibility of the sinusoidal oscillations provides important information on the dynamics underlying the phase transfer process. Figure 2d plots the visibility, defined as $(I_{max} - I_{min})/(I_{max} + I_{min})$, as a function of the fluorescence detection time, for which $t$=0 corresponds to the leading edge of the optical pulse. The data points in Fig. 2d are derived from the sinusoidal oscillations of the fluorescence as a function of $\phi_{opt}$, with the fluorescence collected during a fixed time span of 28 ns. For the numerical fit in Fig. 2d, we used the decay times derived from the corresponding time-resolved fluorescence experiments. The only fitting parameter is the visibility at $t$=0.

To understand the decay of the visibility shown in Fig. 2d, we have carried out CPT experiments, for which the NV is initially prepared in |−1>. Figure 3a shows the fluorescence from |$A_2$> measured as a function of the detuning between the two optical fields. The dip in the spectral response corresponds to CPT and occurs when the detuning satisfies the Raman resonant condition. The spectral width of the dip provides an independent measurement of the optical Rabi frequency[38], which agrees with that determined from the power broadening of the dipole optical transition.

The CPT shown in Fig. 3a arises from optical pumping of the NV center into the dark state. The dynamics of this pumping process is revealed in the time-resolved florescence from |$A_2$> shown in Fig. 3b. For reference, we first set the detuning to be 20 MHz off the Raman resonance such that dark states cannot be formed. In this case, a decay time of 450 ns is observed. This decay arises from optical pumping into the $m_s$=0 state due to state mixing induced by residual strain[39]. We then set the detuning to be at the Raman resonance. The time-resolved florescence now features a much faster decay, with a decay time of 31 ns. This decay time, which is a few times the spontaneous emission lifetime of |$A_2$>, corresponds to the optical pumping of the NV center into the dark state and sets the timescale for the formation of

the dark state. In this regard, Fig. 2d confirms that the decay of the visibility is determined by this optical pumping process, which is much faster than the spin decoherence time.

The maximum fringe visibility observed in Fig. 2 is limited by the fact that the dark state is not completely "dark" due to spin decoherence. In this regard, the maximum visibility is closely related to the depth of the CPT dip in Fig. 3a. We can further improve the visibility by reducing spin dephasing, for example, using isotopically-purified diamond or dressed spin states.

We now turn to the process that encodes the relative optical phase into the superposition spin state and then retrieves the phase with microwave fields. In this case, two Raman resonant optical fields with equal Rabi frequency and a well-defined relative phase prepare an optical dark state via the optical pumping process discussed above. Here, the optical pumping plays the crucial role in driving the NV center into a dark state, which has a relative phase set by $\phi_{opt}$ (see Eq. 2a). This is in contrast to the detrimental role of optical pumping for the optical readout process. As indicated in Fig. 4a, we use two microwave fields with equal Rabi frequency and a well-defined relative phase $\varphi_{mw}$ to drive the electron in the optically-dark superposition spin state to the $m_s$=0 state and then detect the population in the $m_s$=0 state via the resonant $E_y$ transition[17,18]. The duration of the two microwave fields is set to $\tau = \pi / \sqrt{2} \, \overline{\Omega}$, where $\overline{\Omega}$ is the Rabi frequency for the individual microwave fields.

When $\varphi_{mw}$ is in phase with $\phi_{opt}$, the superposition spin state is not only a dark state for the optical fields, but also a dark state for the microwave fields. In this case, the electron spin is also decoupled from the microwave fields, preventing the transfer of the electron population to the $m_s$=0 state. This leads to a minimum in the fluorescence from the $E_y$ state. For comparison, in the limit that $\varphi_{mw}$ is π out of phase with $\phi_{opt}$, the superposition state becomes a bright state for the microwave fields. In this case, the pair of microwave fields serves as a π-pulse, driving the electron from the superposition spin state to the $m_s$=0 state and thus leading to a maximum in the measured fluorescence. Figure 4b shows the sinusoidal oscillation of the fluorescence from the $E_y$ state as a function of $\varphi_{mw}$. The visibility derived from experiments similar to Fig. 4b are plotted in Fig. 4c as a function of the delay between the optical-encoding and microwave-readout pulses. The relatively long decay observed in Fig. 4c is primarily due to the spin dephasing induced by the fluctuating nuclear spin environment. The decay time obtained (0.6 μs) is in agreement with the expected spin dephasing time ($T_2^*$). Note that because of the long spin lifetime, microwave fields are not effective in pumping the electron into a microwave dark state.

The visibility for the microwave readout shown in Fig. 4 is limited to a maximum of 1/5 because of the hyperfine coupling the NV electron spin to the nitrogen nuclear spin. Since the microwave fields used in our experiments were resonant to only one set of hyperfine states and no additional nuclear spin preparation was performed, on average 2/3 of the electron population remains in the $m_s$=0 state and does not participate in the transfer of phase information. Greater fringe contrast can be achieved with nuclear spin preparation[40].

The dark-state based coherent coupling between microwave and optical fields demonstrated above can in principle enable the full quantum state transfer between optical and microwave fields, if an ensemble of NV centers, instead of a single NV center, is used. In this case, the quantum state mapping takes place via a propagating dark state or a dark-state polariton [41]. While there are still considerable technical challenges for using an ensemble of NV centers including inhomogeneous broadening of the NV centers, it is promising to develop a spin-based interface that can enable superconducting circuits to function in a quantum internet.

This work is supported by NSF under grant No. 1604167 and by AFOSR.

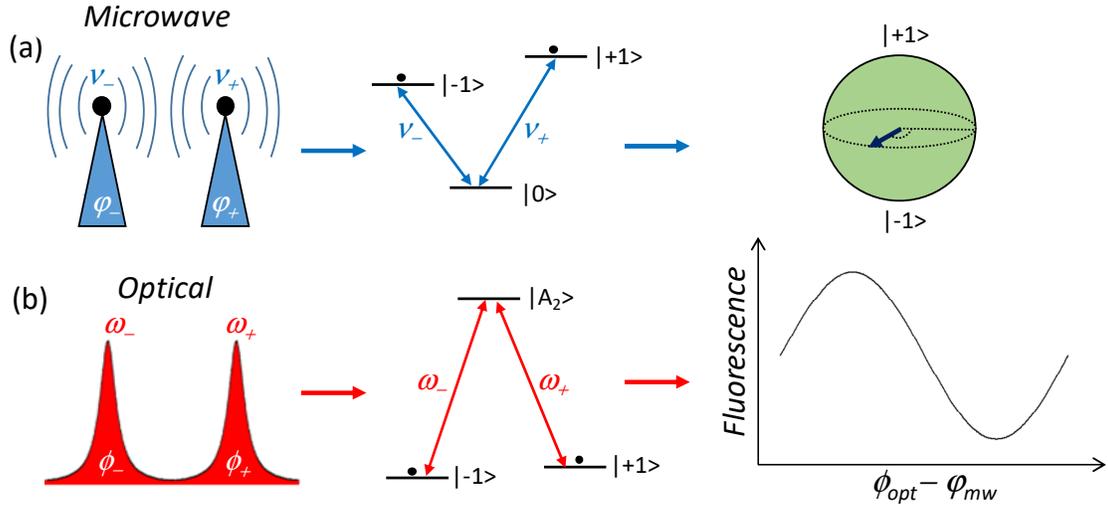

**FIG. 1.** (Color online) (a) Two microwave fields, with relative phase, $\varphi_{mw} = \varphi_+ - \varphi_-$, couple resonantly to the $m_s = 0$ to $m_s = \pm 1$ transitions and generate a coherent superposition of the $m_s = \pm 1$ states, effectively mapping $\varphi_{mw}$ to the phase of the spin coherence. (b) Two optical fields, with relative phase, $\phi_{opt} = \phi_+ - \phi_-$, couple resonantly to the $m_s = \pm 1$ to $A_2$ transitions. The fluorescence from the $A_2$ state features sinusoidal oscillations as a function of $\phi_{opt} - \varphi_{mw}$, effectively reading out the relative microwave phase and transferring the phase information from the microwave to the optical fields.

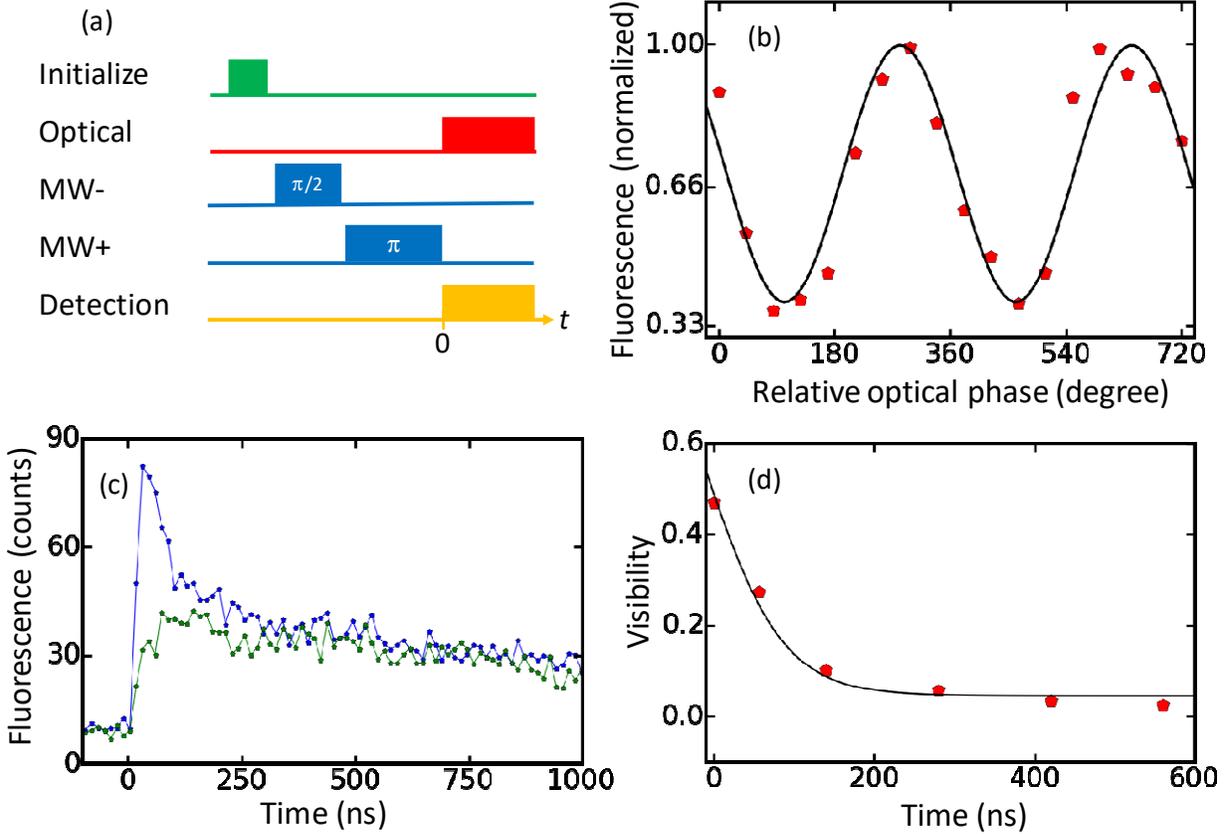

**FIG. 2.** (Color online) (a) Pulse sequence for the transfer of phase information from microwave to optical fields. A NV is initialized in the $m_s$=0 state with an 8 µs green laser pulse. The two MW pulses prepare the NV in a superposition of $m_s$=±1 states. The Raman-resonant optical pulse pair then retrieves the phase information of the superposition spin state. (b) Fluorescence from state $A_2$ detected during the initial 28 ns as a function of relative optical phase. The solid curve is a least-square fit to a sinusoidal oscillation with period $2\pi$, with the peak normalized to 1. (c) Fluorescence from state $A_2$ as a function of time. The upper (lower) curves are obtained with $\phi_{opt}$ near a maximum (minimum) of the oscillations in (b). (d) Visibility of the sinusoidal oscillations as a function of fluorescence detection time, derived from experiments similar to those shown in (b). The solid line shows the results of a numerical fit using the decay time derived from the time-resolved fluorescences.

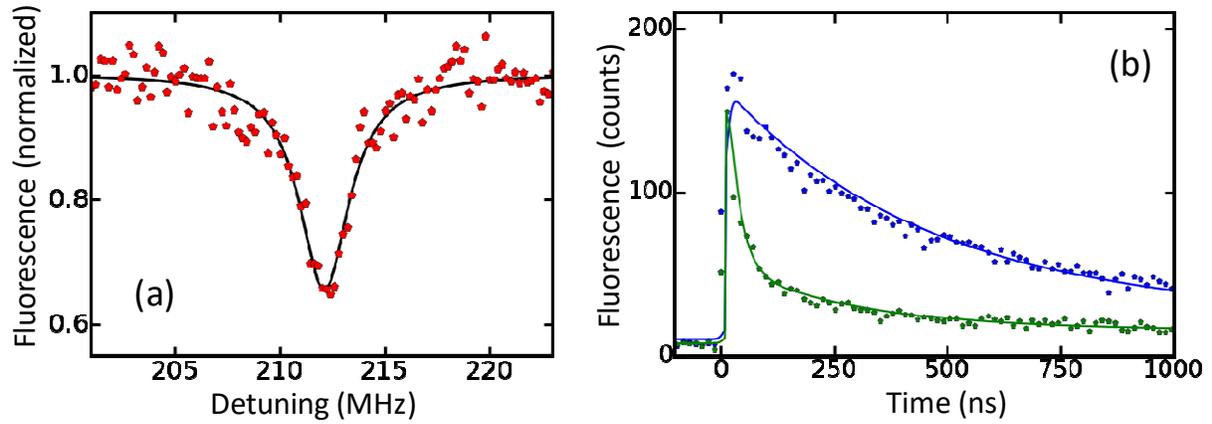

**FIG. 3.** (Color online) (a) Fluorescence from state $A_2$ as a function of the detuning between two applied optical fields, showing the spectral response for CPT. The solid line shows a least-square fit of the dip to an inverted Lorentzian, with a linewidth of 2.8 MHz and with the baseline normalized to 1. (b) Fluorescence from $A_2$ as a function of time. The upper (lower) curve is obtained when the optical detuning satisfies (is 20 MHz away from) the Raman resonance. The solid lines show least-square fits to exponentials, with a decay time of 450 ns for the upper curve. Two decay times, 31 ns and 450 ns, are used for the lower curve.

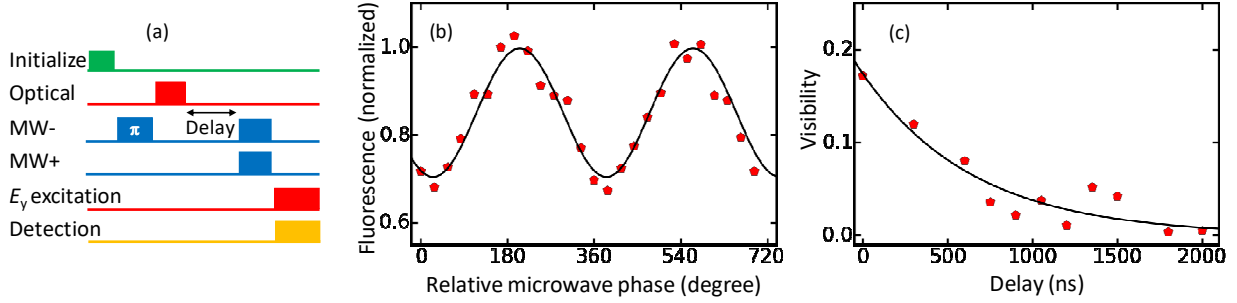

**FIG. 4.** (Color online) (a) Pulse sequence for the transfer of phase information from optical to microwave fields. Following an initialization green laser pulse, a MW $\pi$-pulse puts the NV in the $m_s$=-1 state. The Raman resonant optical pulse pair then prepares the NV in a dark state. A pair of MW pulses read out the phase of the dark state by driving the NV from the dark state to the $m_s$=0 state. Population in state $m_s$=0 is detected via a resonant excitation to the $E_y$ excited state. (b) Fluorescence from state $E_y$ as a function of relative microwave phase. The solid curve is a least-square fit to a sinusoidal oscillation with period $2\pi$ and with the peak normalized to 1. (c) Visibility of the sinusoidal oscillations as a function of the delay between optical-encoding and microwave-readout pulses. The solid line shows a least-square fit to an exponential with a decay time of 0.6 $\mu$s.